\documentclass{aa}
\usepackage[varg]{txfonts}
\bibpunct{(}{)}{;}{a}{}{,}

\begin{document}

\title{\bf  Is a symmetric matter-antimatter  universe excluded?}
\author{J. Baur \inst{1,2} \and A. Blanchard \inst{1} and P. Von Ballmoos \inst{1}}
\institute{IRAP, Observatoire Midi-Pyr\'en\'ees, 14 Av. E.Belin, 31400 Toulouse, France \and CEA, Centre de Saclay, IRFU/SPP,  F-91191 Gif-sur-Yvette, France}

\date{Received xx; accepted xx}

\abstract{ We consider a non-standard cosmological model in which the universe contains as much matter as antimatter on large scales and presents a local baryon asymmetry. A key ingredient in our  approach is that the baryon density distribution follows Gaussian fluctuations around a null value $\eta = 0$. Spatial domains featuring a positive (\textit{resp.} negative) baryonic density value constitute regions dominated by matter (\textit{resp.} antimatter). At the domains' annihilation interface, the typical density is going smoothly to zero, rather than following an abrupt step as assumed in previous symetric matter-antimatter models. As a consequence, the Cosmic Diffuse Gamma Background produced by annihilation is drastically reduced, allowing to easily pass COMPTEL's measurements limits. Similarly the Compton $y$ distorsion and CMB ``ribbons'' are lowered by an appreciable factor. Therefore this model essentially escape previous constrainst on symetric matter-antimatter  models. However, we produce an estimation of the CMB temperature fluctuations that would result from this model and confront it to data acquired from the Planck satellite. We construct a  angular power spectrum in $\delta T / T_{CMB}$ assuming is can be approximated as an average of $C_\ell$ over a Gaussian distribution of $\Omega_B$ using Lewis \& Challinor's CAMB software. The resulting $C_\ell$ are qualitatively satisfying. We quantify the goodness of fit using a simple $\chi^2$ test. We consider two distinct scenarios in which the fluctuations on $\Omega_B$ are compensated by fluctuations on $\Omega_{CDM}$  to assure a spatially flat $\Omega_\kappa = 0$ universe or not. In both cases, out best fit have $\Delta \chi^2 \gtrsim 2400$ (with respect to a fiducial $\Lambda$CDM model), empirically excluding our model by several tens of standard deviations. 
}


\maketitle
\flushbottom

\section{Introduction}
\label{sec:intro}
In the framework of the standard model of particle physics, particles and their corresponding antiparticles are essentially mirror entities. The ringing endorsement of the Big Bang picture, 
whose ancestor is the Lema\^{\i}tre primeval atom, relies on the conjecture that the very early state of the Universe was that of a high temperature plasma in which particles and anti-particles were in thermal equilibrium. Under strict invariance of conjugaison of charge (C), parity (P) and time reversal (T) the Universe would be essentially symmetric in terms of matter and antimatter contents. 

Indeed, in the simplest hot big bang picture, baryons and antibaryons annihilated up until their \textit{freeze-out} from the thermal background, which occurred at $T_B \sim 0.7$ MeV, leaving typical abundances of
$$ \frac{n_B}{n_\gamma}=  \frac{n_{\overline{B}}}{n_\gamma}\sim 10^{-18}$$ 
where $n_{B}$ and $n_{\overline{B}}$ are the number densities of baryons and antibaryons, and ${n_\gamma}$ is the number density of  photons ($\sim 400\:\gamma$.cm$^{-3}$ measured today). These predictions are at odds with conclusions from Big Bang nucleosynthesis (BBN) as the measured abundances of light elements (mainly $^2$H, $^3$He, $^7$Li) are consistent with 
$$ \eta = \frac{n_B}{n_\gamma} \sim 6. 10^{-10}$$
This value of the baryon--to--photon  ratio $\eta$ is completely consistent with the one  derived from the cosmic background radiation (CBR) fluctuations. \\

Early attempts at building physically-motivated matter-antimatter symmetric universes \citep{1973A&A....28..253A} encountered serious difficulties~\citep{1975Natur.253...25C,1976ARA&A..14..339S}. Furthermore, the lack of any evidence for the presence of antimatter has surged support for the standard picture, in which a matter-antimatter asymmetry is generated through a mechanism beyond known standard physics. Indeed, the discovery of the CP (charge-parity) violation in accelerators has rendered possible an early creation of a significant baryon-antibaryon asymmetry called baryogenesis. Sakharov identified in the early 1970s the three criteria required for baryogenesis: violation of both B (baryon charge) and CP, and a departure from thermal equilibrium. Although these conditions are met in the standard model, no noticeable asymmetry can be generated at energies currently accessible in present-day particle accelerators. We are therefore left to conjecture that the mechanism responsible for baryogenesis lies beyond the scope of the standard model of particle physics, for instance during the reheating phase following inflation. Although such mechanisms could ensure an asymmetry, whether they lead to a global or a local one remains open. In the latter case, local asymmetry refers to spatial fluctuations of the baryon density resulting in a patchwork of distinct regions alternatively dominated by matter or antimatter, all-the-while ensuring a global neutrality \citep{1979PhRvL..43..315B}. Such models are usually refered to as Local Asymmetry Domain Cosmologies~\citep{2002hep.ph....7323S}, LADC hereafter. In this work, we further investigate this possibility by assuming that the baryon asymmetry fluctuates {\em according to matter fluctuations}, leading to a fluctuating {\em baryon field} $\delta_B$:
$$\delta_B \propto n_B   \propto \frac{\delta \rho} {\rho}  $$
with $n_B$ encoding $-n_{\overline{B}}$ in antimatter-dominated regions. However, the CDM matter power spectrum presents fluctuations on every scale down to very small ones. Therefore the above relation cannot hold as it is robustly established that domains of matter and antimatter should be at least 20 Mpc in size \citep{1976ARA&A..14..339S,1996NuPhS..48..514D}. In what follows, we thus assume that the power spectrum of the baryon asymmetry field $\delta_B$ cuts-off at some scale $\lambda_{\rm{mam}} \ge 20$ Mpc. In that respect, our assumption is reminescent of the standard patchwork-universe picture. However, a differing assumption we are making on the geometry of matter (antimatter) domains allows one to circumvent the usual constraints on patchwork LADC models for the reasons detailed hereafter.\\


\section{Physics of matter antimatter interface}
\label{sec:PMI}
Differences between standard and LADC models primarily arise from the presence of interface regions between domains where matter and antimatter meet and annihilate. The physics of this annihilation zone has been the subject of several investigations in the past \citep[CDG hereafter]{1973A&A....28..253A,1976ARA&A..14..339S,1998ApJ...495..539C}. The characteristic thickness $\lambda_A$ of the annihilation zone can be roughly estimated as the typical thermal-travel distance of a proton assuming no supersonic flow in matter~:
$$ \lambda_A(z) \approx \frac{v_{th}(z)}{H(z)}$$
At high redshifts ($z \ge 300$) the CBR drag on the fluid lowers this value. Because of annihilation and hydrodynamics of the fluid, there is a depletion zone of typical thickness $\lambda_D$ wherein the baryon fluid density drops considerably. $\lambda_D$ is anticipated to be of the same order of the sound-travel distance and therefore of the same magnitude as $\lambda_A$ at low redshifts. Products of annihilation include high-energy electrons that lose their energy as they propagate, warming up the intergalactic medium over a typical length scale $\lambda_H$. Within this region, the intergalactic medium is heated to temperatures varying from a few to tens of eV~\citep{1998ApJ...495..539C}. 

\subsection{The resulting diffuse gamma-ray background}

The origin of the MeV gamma-ray background is presently unclear \citep{2015arXiv150206116R}. A solution proposed in the past was that it is generated by matter-antimatter annihilation at cosmological scales assuming a LADC. The background flux is the integral of the number of annihilation events $J$ per unit time and unit surface area delimiting the annihilation interfaces:
\begin{equation}
J \; \doteq \; \int \langle \sigma_{ann} v \rangle n_p \bar{n}_p  dl
\end{equation}
In an infinitely thin annihilation zone approximation, the annihilation rate per unit area orthogonal to the interface becomes:
\begin{equation}
J \sim n_p  v_{th}(z) 
\end{equation}
 In their seminal paper, CDG provide extensive calculations of annihilation processes. Their results can be used to infer predictions for our model by scaling arguments. A central ingredient in their model is that the number density of baryons (or antibaryons) beyond the depletion region (of typical size $\lambda_D$) rapidly reaches the universal value $n_B$.  The numerical value of the comoving $\lambda_D^c$ can be evaluated roughly as the sound-travel distance. Assuming a typical temperature of $3 \cdot 10^3$K for the medium, one gets (with $\Omega_M \sim 0.3$): 
\begin{equation}
\lambda_D^c \approx (1+z)\frac{v_{th}}{H(z)} \approx 10 \left(\frac{300}{1+z}\right)^{1/2} h^{-1} \textrm{kpc}
\end{equation}
In our model, the baryon number density varies smoothly with position over the extension of the patch from its typical value in the center of the domain $n_B$ to a much smaller value $ n $ close to the interface. We can estimate the average density $n$ in this region by:
\begin{equation}
n \approx \frac{\lambda_D^c}{\lambda_{\rm{mam}}^c} n_B \le 10^{-3}n_B
\label{eq:red}
\end{equation}
Such a drastic decrease in the density at the annihilation interface easily reduces the diffuse gamma-ray background (DGB) below observational limits established by COMPTEL \citep{comptel,comptel2}. The DGB constraint of CDG is no longer problematic as soon as domains are larger than a typical size of 20 $h^{-1}$Mpc.
This is an important conclusion, as the argument of the CDG paper is generally regarded as the strongest evidence against a patchwork universe in which the typical size of domains is smaller than the present-day horizon \citep{2012NJPh...14i5012C}.

\subsection{Other probes or constraints on LDAC}

Here we briefly discuss two aspects that have been previously adressed in past litterature. The energy carried out by electrons produced by annihilation is transferred to CBR photons via Compton scattering, producing a Compton distorsion of the spectrum whose amplitude is measured by the $y$ parameter~\citep{SZ}. CDG estimated that the expected amplitude of $y$ in their patchwork LADC would be below the stringent limit established by the FIRAS instrument~\citep{FIRAS}. The $y$ distorsion parameter being proportional to $J$, its numerical value is reduced by the same factor in Eq.\ref{eq:red} above in our model. Another consequence is the possible presence of ribbon-like imprints on the CBR through the $y$ distorsion produced by the matter-antimatter interface \citep{1997PhRvL..79.2620K}. However,  \citet{1998ApJ...496L..63C} revised the previous estimate and found an appreciably lower amplitude for the predicted signal, beyond sensitivity of space missions like Planck. In our model the signal is again reduced by the same factor in Eq.\ref{eq:red}, leaving no hope for detecting such a signal. \\

A more serious concern about LADC models comes from the vantage point of structure formation. One may hope that once structures are formed, matter and antimatter domains are sufficiently separated to avoid devastating annihilation. The absence of significant gamma-ray emission from clusters, including colliding clusters, can be used to infer constraints on the domains' characteristic size \citep{2008JCAP...10..001S,2015PhRvD..91h3002P}. However with our assumption, regions of cluster formation correspond to positive overdensities ($\delta \rho/\rho \ge \delta_{\rm{thres}}$) and are therefore expected to have a pure baryon component devoid of any antimatter. This is likely to hold as long as the smoothing length 
is of the same order of the initial scale of large clusters, which is of the order of 20 $h^{-1}$Mpc. It remains to be checked whether a scenario of structure formation could be built on {\em smaller scales}: from time to time, galaxy-sized halos should form in regions overlapping an interface region. This is due to the fact that correlations between small and large scales should wash out as the difference in scales increases. In such a case, the halo would contain significantly similar amounts of matter and antimatter, leading to pronounced annihilation, thus potentially reviving the overproduction of the gamma diffuse background.


\section{Constraints from the Microwave Background}
\label{sec:CMB}
The temperature fluctuations over the celestial sphere are usually expressed by using  spherical harmonic decomposition :
\begin{equation}
\label{eq:Alm}
\frac{\delta T}{T} \left( \theta,\varphi \right) = \sum_{\ell=0}^\infty \sum_{m=-\ell}^{+\ell} a_\ell^m \mathcal{Y}_\ell^m \left( \theta, \varphi \right)
\end{equation} 
where $\ell$ is the multipolar moment, $m$ the mode of degeneracy and the $\mathcal{Y}_\ell^m$  are the spherical harmonics.
The angular power spectrum of the CMB temperature fluctuations is defined as:
\begin{equation}
\label{eq:Cell}
C_\ell = 
\frac{1}{2 \ell + 1} \sum_{m = - \ell}^{+ \ell} |a_\ell^m|^2
\end{equation}

Currently, the most constraining data for cosmological models is the measurement of the angular power spectrum of the cosmic microwave background (CMB) fluctuations $\mathcal{C}_\ell$~\citep{2015arXiv150201589P}. In the following, we use Planck TT CMB data to constrain our LADC model. The full calculation of the TT angular power spectrum is typically done through integration of a dedicated Boltzmann code. Here, we propose an approximation for computing the expected $\mathcal{C}_\ell$ in the scope of our LADC model without going through the full extent of such a code.  

The characteristic size of domains in our model is of order $\gtrsim 20 \:h^{-1}$Mpc, which is the typical thickness of the region wherein photons from the last scatering surface are originating. It is worthy to note that this length is insensitive to the value of baryon content. Consequently, when looking at the CMB, one essentially views independent regions of space which do not overlap along the line-of-sight. Each region is a realization of the Gaussian field generating the baryon field. Therefore the baryon density in units of critical density $\omega_B = h^2 \Omega_B$ averaged over a domain is a semi-normally-distributed random variable :
\begin{equation}
\label{eq:distrib}
p_{\sigma_B} (\omega_B) = \frac{2}{\sqrt{2 \pi} \sigma_B} \exp \left[ - \frac{1}{2} \left( \frac{\omega_B}{\sigma_B} \right)^2 \right]
\end{equation}
where $\omega_B$ denotes $-\omega_{\bar{B}}$ in antimatter-dominated domains, since baryons and antibaryons behave alike with regards to CBR photons. 
An intrinsic free parameter in our model is the (semi-normal) standard
deviation of baryon density $\sigma_B$. There are two scale ranges to consider while estimating the fluctuations spectrum in our LADC model according to whether  modes  are lesser or greater than those corresponding to $\lambda_{\rm{mam}}$, \textit{i.e.}, $\ell \sim  \pi /\theta_{\rm{mam}} \sim \pi /(\lambda_{\rm{mam}} D_{A}^{\rm{CMB}})$ where $D_{A}^{\rm{CMB}}$ denotes the comoving angular distance to the last scatering surface. In the latter case, we can expect that within a single patch, the physics determining the $\mathcal{C}_\ell$ is that of a cosmological model with a roughly constant local value of $\omega_B$ (in absolute value as it is the same physics for matter and antimatter). This means that over a patch denoted by the subscript $i$ covering a solid angle $\Omega_i$ over the sky, one experiences a fluctuation field $\delta T/T|_i$ featuring a correlation function  $\xi_i(\theta)$ for scales $\theta$ smaller than $\theta_{\rm{mam}}$ that is essentially that of a cosmological model with the baryon density $\omega_B$, and the corresponding angular power spectrum for patch $i$ is therefore (for $\ell \gtrsim \frac{\pi}{\theta_{\rm{mam}}}$):
\begin{equation}
\mathcal{C}_\ell(\omega_B) \frac{\Omega_i}{4 \pi}
\end{equation}
For small angular scales, different patches represent distinct and independent realizations of the field, and so the total angular power spectrum is:
\begin{equation}
\label{eq:meanCell}
\mathcal{C}_\ell \approx \sum \mathcal{C}_\ell(\omega_B) \frac{\Omega_i}{4 \pi} \approx \langle \mathcal{C}_\ell(\omega_B) \rangle
\end{equation}
where the average is taken over the different patches, {\em i.e.} over the realizations of $\omega_B$ given by the distribution in Eq.\ref{eq:distrib}.

 
For scales larger than $\theta_{\rm{mam}}$ we apply the same reasoning~: for scales associated to some $\ell$, the fluctuations are coming from different regions of scale  $\pi /\ell$ over the sky which are essentially independent and we should therefore expect the same relation as in Eq.\ref{eq:meanCell}, which consequently holds in all cases. We use the \textsf{CAMB} software\footnote{\tt http://camb.info}~\citep{Lewis2000} to produce a power spectrum $C_\ell (\omega_B)$ for each value of our $\omega_B$ sample and average them to produce a mean power spectrum defined in Eq.\ref{eq:meanPS} :
\begin{equation}
\label{eq:meanPS}
\langle C_\ell^{\sigma_B} \rangle = \frac{\sum p_{\sigma_B}(\omega_B) \cdot C_\ell (\omega_B)}{\sum p_{\sigma_B}(\omega_B)} 
\end{equation} 
The weights $p_{\sigma_B}(\omega_B)$ in Eq.\ref{eq:meanPS} are sampled from Eq.\ref{eq:distrib}.
We sample $\sigma_B$ in the range $ \left[ 3, 53 \right] \times 10^{-3}$.
We assume a spatially flat Universe and consider two cases: in the first one, we assume baryon fluctuations to be compensated by 
cold dark matter fluctuations, in the second case we assume the cold dark matter density $\Omega_C$ to stay constant. We refer to the former as the \textit{constant $\Omega_M$} case and the latter as the \textit{constant $\Omega_C$}. Apart from these, all other input parameters of \textsf{CAMB} remain fixed at their fiducial values summarized in Tab.1.\\

\begin{table}[htb]
\begin{center}
\begin{tabular}{lc}
\textbf{input parameter} & \textbf{central value}\\[2pt]
\hline \\[-10pt]
$h$ &  $0.6731$\\[2pt]
$n_s$ &  $0.9655$\\[2pt]
$\Omega_B h^2$ &  $0.02222$\\[2pt]
$\Omega_C h^2$ &  $0.1197$\\[2pt]
$\Omega_\Lambda$ &  $0.685$\\[2pt]
$100 \Omega_\kappa$ &  $0.0$\\[2pt]
$\Omega_\nu h^2$ &  $0.000644$\\[2pt]
$N_{\rm{eff}}$ &  $3.046$\\[2pt]
\end{tabular}
\end{center}
\label{tab:params}
\caption{Set of \textsf{CAMB} input parameters for our $\Lambda$CDM fiducial model. Cosmological values are based on Planck 2015 \citep{2015arXiv150201589P} best fit using TT+lowP data set. Massless neutrino fraction is set to 2.046 and massive neutrino fraction to 1. $\Omega_\nu$ is obtained by setting $\sum m_\nu = 60 \: \rm{meV}$.}
\end{table}

Note that in the constant $\Omega_C$ case, the value of $\Omega_M$ varies locally according to the local variations in $\Omega_B$. Therefore, to compute $C_\ell$ in such cases, we use \textsf{CAMB} for the flat
 model  using the modified value of $\Omega_M$ and correct for the (irrelevant) change in the  angular diameter distance $D_A$ at $z=1050$.  From each $C_\ell^{\rm{CAMB}}$ produced by \textsf{CAMB} for a given value of $\omega_B$, we issue a corrected $C_{\ell^\star} (\omega_B)$ where $\ell^\star$ are the corresponding values of the multipole in a fixed $\Omega_M$ cosmology :

\begin{equation}
\label{correctionEll}
\ell^\star (\omega_B) = \ell \times \frac{D_A (\Omega_M (\omega_B), z=1050)}{D_A(\Omega_{M0}, z=1050)}
\end{equation}
The above Eq.\ref{correctionEll} only applies to the constant $\Omega_C$ case, in which $\Omega_M$ and thus $\ell^\star$ vary as functions of our $\omega_B$ sample. In the constant $\Omega_M$ case, fluctuations in $\omega_B$ are compensated by fluctuations in $\Omega_C$, to which the angular diameter distance is independent of. We apply this correction prior to averaging the power spectrum in Eq.\ref{eq:meanPS}. For simplicity sake, we drop the $\ell^\star$ notation in these equations since it has heretofore been accounted for variations in $D_A (\Omega_M, z=1050)$. \\ 

We define the optimal mean power spectrum for our model the $\langle \mathcal{D}_\ell^{\sigma_B} \rangle = \langle C^{\sigma_B}_\ell \rangle \times \ell (\ell+1)/2 \pi$ that minimizes the quantity:
\begin{equation}
\label{eq:chi2}
\chi^2 \left( \Upsilon, \sigma_B \right) = \sum_\ell \left( \frac{\Upsilon \langle \mathcal{D}^{\sigma_B}_\ell \rangle - \mathcal{D}_\ell^\star}{\sigma^\star_\ell} \right)^2
\end{equation} where $\mathcal{D}_\ell^\star$ and $\sigma_\ell^\star$ are respectively the value and standard deviation at multipole $\ell$ of the TT angular power spectrum from the Planck collaboration. The amplitude of the fluctuation is let free through the $\Upsilon$ constant. Our rudimentary definition of a $\chi^2$ test given by Eq.\ref{eq:chi2} does not account for multipole correlations and other rigorous statistical analysis the Planck collaboration subjected their study to. We therefore control for the above forementioned $\chi^2_{\rm{min}}$ values by comparing them to the value yielded by inputing the set of parameters that best fits the Planck 2015 TT+lowP data (see Tab.1). We refer to our fiducial model as the $\Lambda \rm{CDM}$ case in Tab.2 below which recaps the optimized parameter and $\chi^2_{\rm{min}}$ values. The $\sigma_B$ values have been converted into their corresponding $\Omega_B h^2$ by computing the expectation values of Eq.\ref{eq:distrib}.\\

\begin{table}[htb]
\begin{center}
\begin{tabular}{cccc}
\textbf{ } & \textbf{$\Omega_M = \rm cst$} & \textbf{$\Omega_C = \rm cst$} & \textbf{$\Lambda \rm CDM$}\\[2pt]
\hline \\[-10pt]
$\chi^2_{\rm{min}}$ &  $3425$ & $3465$ & $1031$ \\[2pt]
$\Upsilon$ &  $1.101$ & $1.095$ & $1.019$ \\[2pt]
$\langle | \omega_B | \rangle$ &  $0.0191$ & $0.0211$ & $0.0222$ \\[2pt]
\end{tabular}
\end{center}
\label{tab:results}
\caption{Set of optimal $\left( \Upsilon, \langle |\omega_B| \rangle \right)$ parameters for minimizing Eq.\ref{eq:chi2} for both constant $\Omega_{M,C}$ cases and our fiducial $\Lambda \rm CDM$ case. 
}
\end{table}

In both constant $\Omega_{M}$ and $\Omega_{C}$ cases, the optimized power spectrum yield $\Delta \chi^2$ values of $\sim 2,400$. These disparities settle our model as inconsistent with CMB temperature anisotropies data by several tens of standard deviations. 

\begin{figure*}
\centering
\includegraphics[width=16cm]{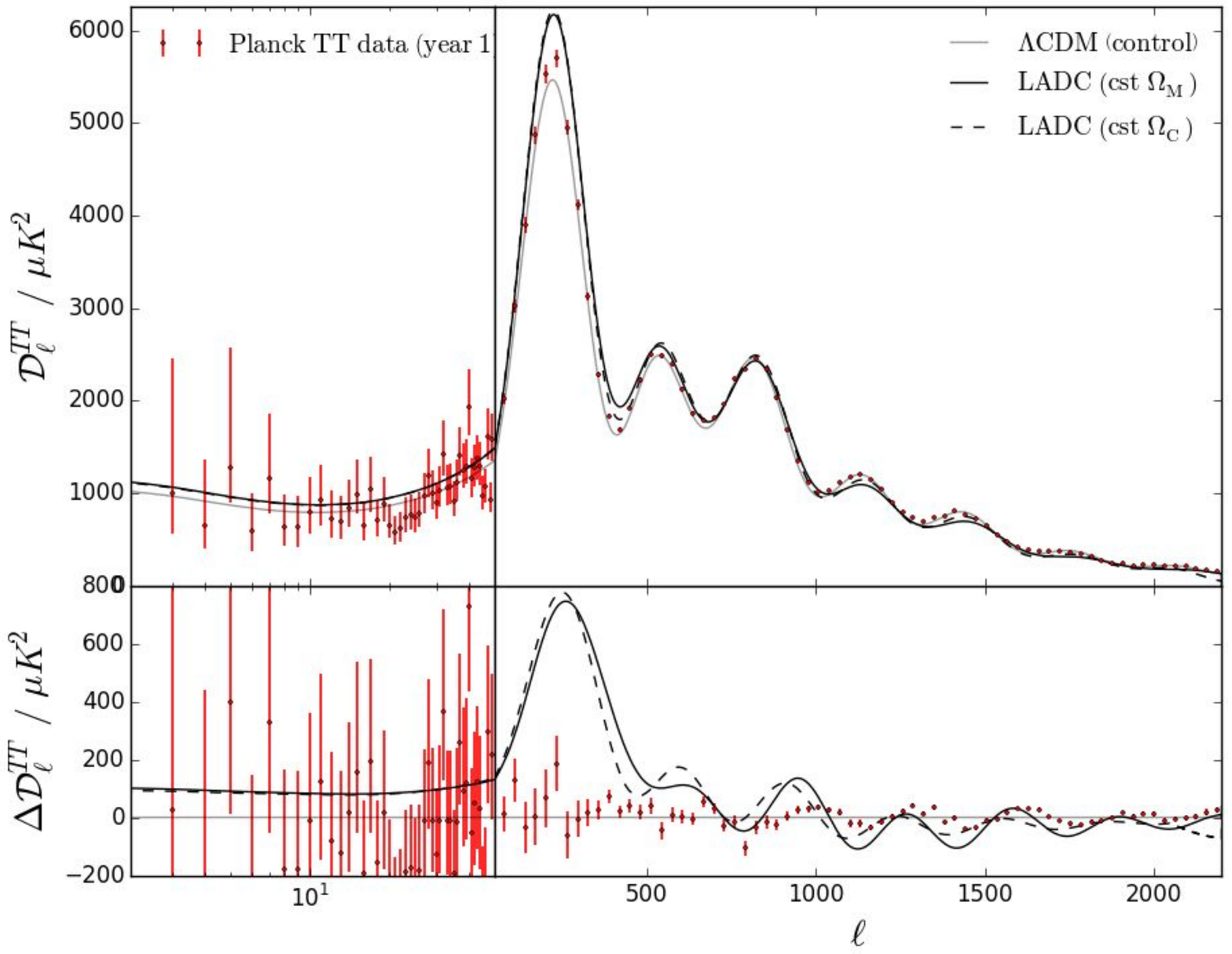}
\caption{\textbf{Top Panel} -- Angular power spectrum coefficients $\Upsilon^\star \mathcal{D}^{\sigma_B^\star}_\ell$ obtained with the optimized values in Tab.2 for both constant $\Omega_{M,C}$ cases (solid and dashed black lines, \textit{resp.}). Our fiducial $\Lambda$CDM model is shown in grey. Red error bars are  {\em Planck} spacecraft data. \textbf{Bottom Panel} -- $\mathcal{D}_\ell$ residual values with respect to our fiducial $\Lambda$CDM model. Angular multipole scale is logarithmic under $\ell < 50$ for sake of visibility.} 
\label{fig:Dell}
\end{figure*}


\section{Summary}
\label{sec:results}
The simplicity of a symmetric matter-antimatter universe, {\em i.e.} a universe containing on average an equal quantity of matter and antimatter, makes such a possibility rather attractive. Indeed, several authors considered this possibility in the recent past despite its rejection based on  the absence of a mechanism for matter-antimatter separation and on the lack of any significant relic flux of diffuse gamma rays.  This last issue is generaly regarded as the most severe drawback of these scenarios. \\

We have introduced a new class of LADC models in which local fluctuations in the baryon density are Gaussian, vary smoothly with position and are correlated with matter fluctuations on large scale ({ \em i.e.} $\delta n_B \propto \delta$). In this model, the density of matter and antimatter is drastically reduced
at  domains interface. The flux of diffuse gamma rays is reduced accordingly and remains well below experimental limits. 
Similarly, Compton $y$ distorsion and induced ``ribbons'' are lowered to unobservable values. Such models therefore essentially escape classical constraints on matter-antimatter models. 
Nevertheless, accurate constraints in modern cosmology are now obtained from the measurements of temperature fluctuations in the CMB. We have therefore estimated  the characteritics of the angular spectrum of the CMB fluctuations $C_\ell$ in the above types of models. 
The average behavior of the $ C_\ell$ is in qualitative good agreement with the measured TT angular power spectrum (see upper panel of Fig.\ref{fig:Dell}). However, a quantitative simple $\chi^2$ analysis shows  that the LADC models proposed here are in serious difficulty in reproducing  the Planck spacecraft data, 
the  models being at odds with a $\Lambda$CDM cosmology by several tens of standard deviations. Despite their simplicity,
our calculations give  clear indication that symmetrical matter-antimatter models are confronted to an additional constraint that will be difficult to bypass. \\



\section*{Acknowledgments}

JB is thankful to the Observatoire Midi-Pyr\'en\'ees and the Institut de Recherche en Astrophysique et Plan\'etologie for their ressources and accomodation during parts of this research, as well as James Rich and Ludovic Montier for useful comments and suggestions for this work.\\

\bibliographystyle{aa}
\bibliography{biblio}
\end{document}